\documentclass{egpubl}
 
 \JournalSubmission    % uncomment for submission to Computer Graphics Forum
\usepackage[T1]{fontenc}
\usepackage{dfadobe}  

\usepackage{wrapfig}
\usepackage{bm}
\usepackage{amsmath}
\newcommand{\eg}{\textit{e.g.}, }
\newcommand{\ie}{\textit{i.e.}, }
\newcommand{\etal}{\textit{et al.}}
\newcommand{\one}[0] {{\em (i)}~}
\newcommand{\two}[0] {{\em (ii)}~}
\newcommand{\three}[0] {{\em (iii)}~}

\DeclareMathOperator*{\argmax}{argmax}

\biberVersion
\BibtexOrBiblatex
\usepackage[backend=biber,bibstyle=EG,citestyle=alphabetic,backref=true]{biblatex} 
\electronicVersion
\PrintedOrElectronic

\ifpdf \usepackage[pdftex]{graphicx} \pdfcompresslevel=9
\else \usepackage[dvips]{graphicx} \fi

\usepackage{egweblnk} 
\title[Curvy]%
      {Curvy: An Interactive Design Tool for Varying Density Support Structures}

\author[E. Ulu, N. Gecer Ulu, J. Li \& W. Hsiao]
{\parbox{\textwidth}{\centering E. Ulu$^{1}$\orcid{0000-0002-1612-3123}
        N. Gecer Ulu$^{1}$\orcid{0000-0002-5164-387X} 
        J. Li$^2$
        and W. Hsiao$^1$
        }
        \\
{\parbox{\textwidth}{\centering $^1$Palo Alto Research Center, USA\\
         $^2$UCLA HCI Research, USA
       }
}
}
\begin{document}

\teaser{
 \includegraphics[width=\textwidth]{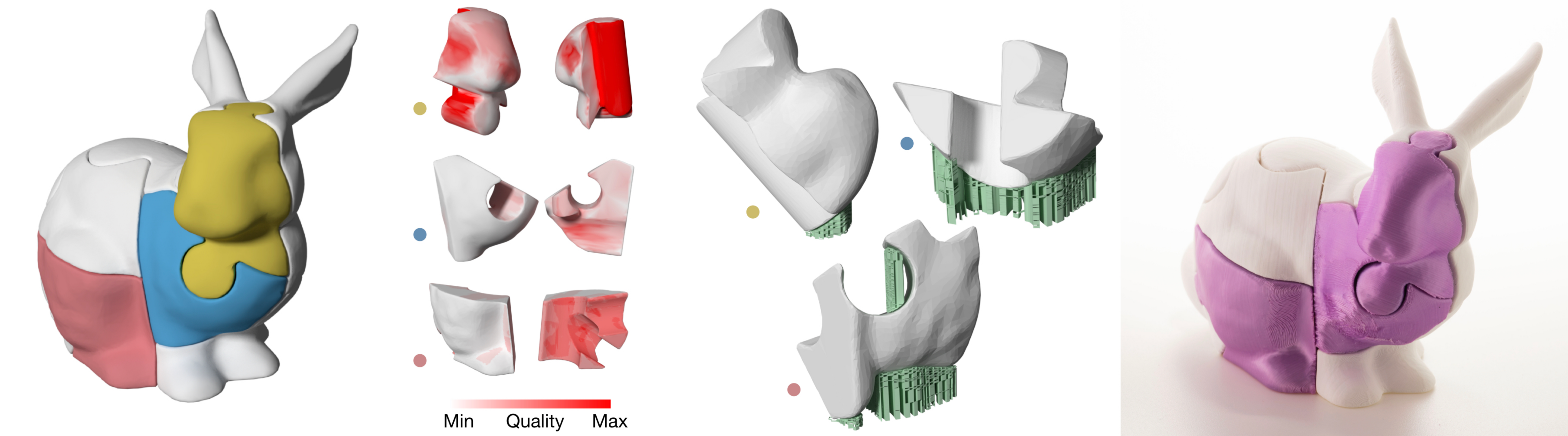}
 \centering
  \caption{Curvy enables users to design support structures implicitly, by providing high-level surface quality preferences (middle-left) directly on the target object (left). User preferences are translated into low-level support parameters to generate varying density curvy zigzag supports (middle-right). Resulting prints satisfy perceptual and functional intents of the user (right). Puzzle pieces fit together well while visually important surfaces are free of support contact marks. In the middle figures, puzzle pieces are marked with colors corresponding to the ones on the left model.}
\label{fig:teaser}
}

\maketitle
%-------------------------------------------------------------------------
\begin{abstract}
We introduce Curvy--an interactive design tool to generate varying density support structures for 3D printing. Support structures are essential for printing models with extreme overhangs. Yet, they often cause defects on contact areas, resulting in poor surface quality. Low-level design of support structures may alleviate such negative effects. However, it is tedious and unintuitive for novice users as it is hard to predict the impact of changes to the support structure on the final printed part. Curvy allows users to define their high-level preferences on the surface quality \textit{directly} on the target object rather than explicitly designing the supports. These preferences are then automatically translated into low-level design parameters to generate the support structure. Underlying novel \textit{curvy} zigzag toolpathing algorithm uses these instructions to generate varying density supports by altering the spacing between individual paths in order to achieve prescribed quality. Combined with the build orientation optimization, Curvy provides a practical solution to the design of support structures with minimal perceptual or functional impact on the target part to be printed.
%-------------------------------------------------------------------------
%  ACM CCS 1998
%  (see https://www.acm.org/publications/computing-classification-system/1998)
% \begin{classification} % according to https://www.acm.org/publications/computing-classification-system/1998
% \CCScat{Computer Graphics}{I.3.3}{Picture/Image Generation}{Line and curve generation}
% \end{classification}
%-------------------------------------------------------------------------
%  ACM CCS 2012
  %(see https://www.acm.org/publications/class-2012)
%The tool at \url{http://dl.acm.org/ccs.cfm} can be used to generate
% CCS codes.
%Example:


\end{abstract}  
%-------------------------------------------------------------------------
\section{Introduction}

In many 3D printing approaches, support structures play an essential role in successful printing by \textit{supporting} the overhang regions to prevent them from collapsing under gravity. While they may be necessary, these auxiliary structures often damage the surface quality resulting in visual as well as functional artifacts. Such artifacts are often unavoidable especially in widely used single material fused filament fabrication (FFF) process as there is a delicate balance between the amount of supports and the resulting surface quality. When supports are used excessively, they stick to the walls of the models and leave blemishes on the surface. On the other hand, when insufficient, sagging occurs between the support contact points due to the large bridging distance. Many 3D printing software, such as Autodesk MeshMixer \cite{meshmixer}, Simplify3D \cite{simplify3d}, Ultimaker Cura \cite{cura} and Slic3r \cite{slic3r}, provide automated means to design support structures. However, without user involvement, resulting supports often lead to suboptimal surface quality even for simple shapes. User involvement required to alleviate the quality issues, on the other hand, is generally very low-level and tedious, such as manual placement of individual support pillars, local adjustments of support patterns, its orientation and spacing. Although such adjustments are attainable for experienced designers, a more intuitive and direct way of support design is required for novice users.

%\begin{figure}
%    \centering
%    \includegraphics[width=\columnwidth, height=2in]{example-image}
%    \caption{Showing surface quality of a model that is printed  using Meshmixer supports, ZigZag supports from one/a few slicers, etc.}
%    \label{fig:fig1}
%\end{figure}

We present Curvy--an interactive tool to design support structures with minimal perceptual and functional impact on the object. Curvy takes as input users' high-level specifications on the desired surface quality and produces support toolpaths for each layer accordingly (Figure~\ref{fig:teaser}). As the user input is defined \textit{directly} on the surface of the target model rather than the supports, users do not need to have a low-level knowledge on inner workings of the support generation algorithm to predict the impact of changes they are making on the final printed result. 

Main challenge in such a direct approach lies in the translation of the high-level inputs on the target object to the low-level support design parameters. For such a transition to be possible, underlying support parametrization needs to be sufficiently flexible to comply with any arbitrary input whereas general enough to be applicable to variety of shapes. Additionally, resulting supports are required to exhibit common properties including easy removal and low additional cost to print.

Our approach overcomes this challenge using a novel toolpathing approach that \one generates curvy zigzag paths conforming to the shape boundaries as well as the infill patterns of each layer and \two controls the spacing between individual paths (\ie density) locally. The former capability allows us to obtain easy to remove supports while consistently providing sufficient amount of contact points for successful bridging. The latter one, on the other hand, enables local control of the surface quality per user prescriptions on the target object. Combined with a build direction optimization approach, unnecessary supports as well as supports touching the regions that are intended to have high surface quality are minimized.%, thereby reducing print time and material use.

Our main contributions are:
\begin{itemize}
    \item a novel curvy zigzag toolpathing algorithm that results in high surface quality while allowing easy removal,
    \item a method to manipulate support density locally,
    \item an interactive support structure design tool that allows users to define high-level preferences directly on the target object using above two ideas.
\end{itemize}

%-------------------------------------------------------------------------
\section{Related Work}

Curvy aims to provide end-users an interactive way to design support structures implicitly by defining their high-level surface quality preferences directly on the target object. This goal cross-cuts three areas of prior work in 3D printing: \one computational tools for design and process planning, \two support structures generation methods and \three build orientation optimization approaches.

\subsection{Design and Process Planning Tools}
As simple as it may seem, 3D printing is a complex process with many aspects that require low-level design and process planning for a successful operation. In order to make it more accessible to novice users and more convenient to experienced users, past research has explored variety of interactive tools targeting many different aspects of 3D printing. Recent examples include two-piece mold design \cite{nakashima2018corecavity}, modeling through augmented reality \cite{peng2018roma}, patching for minimum waste design iterations \cite{teibrich2015patching} and deformable object design \cite{he2019ondule}. Similar to these approaches, Curvy aims to alleviate the amount of low-level design users need to perform for a successful 3D print. In particular, Curvy focuses on making design of support structures easy and intuitive by eliminating the need for explicit manipulation of support structures. Instead, the support structures are designed implicitly, by specifying the high-level quality preferences directly on the target object.

Other approaches focus on providing optimal slicing to improve the surface quality or geometric accuracy. Recent examples of such approaches include adaptive \cite{wang2015saliency, alexa2017optimal} and curved slicing \cite{etienne2019curviSlicer} schemes. These methods improve the quality of the prints by mainly alleviating the so-called staircase effect. Our approach is complementary to these tools in that presented slicing schemes may be facilitated in printing the target object to further improve the surface quality while Curvy is minimizing the impact of supports on the final result.

\begin{figure*}
    \centering
    \includegraphics[width=\textwidth]{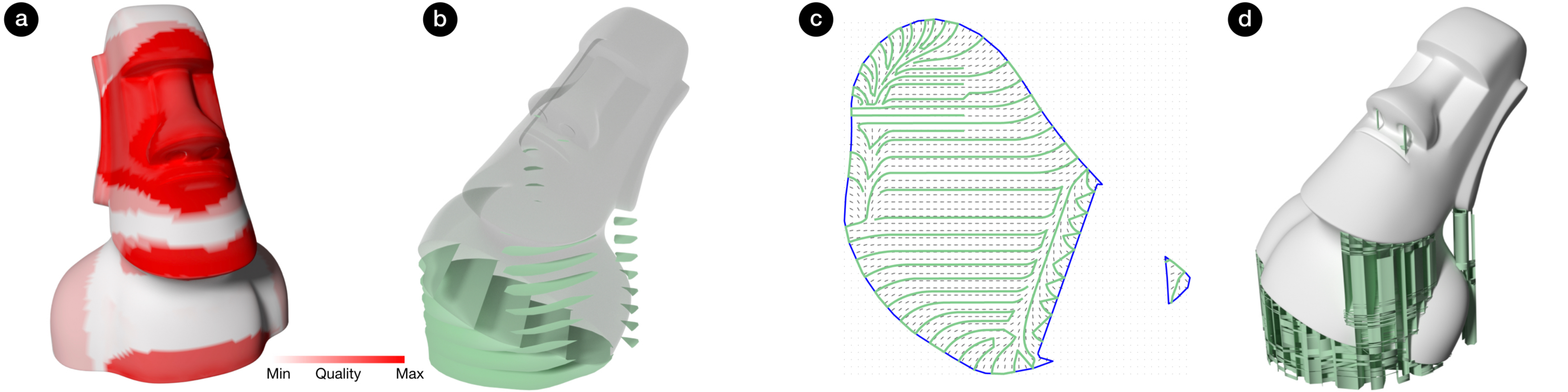}
    \caption{Given an input 3D geometry in an arbitrary initial orientation and users' preferences in desired surface quality (a), Curvy optimizes the build orientation to avoid supports in high quality regions as much as possible (b) and generates variable density curvy zigzags in each support polygon (c) to create a final support structure (d). Green represents support polygons in (b) and generated support toolpaths in (c-d).}
    \label{fig:fig2}
\end{figure*}

\subsection{Support Structure Generation}
Generating supports is often composed of two main steps: \one detection of surfaces requiring supports (\ie overhangs), and \two design of the support structure itself. For detection of overhangs, Chalsani \etal~\cite{chalsani1995support} performed a boolean difference between two successive slices while Kirschman \etal~\cite{kirschman1991computer} and Allen~\etal \cite{allen1995determination} considered down-facing facets of the input mesh having angle too steep to print correctly as overhangs. In our build orientation optimization, we employ the latter as it is computationally more practical while being sufficiently accurate.

While commercial 3D printing software such as Simplify3D \cite{simplify3d}, Ultimaker Cura \cite{cura} and Slic3r \cite{slic3r} provide general purpose space filling support structures in the form of regular patterns such as zigzags or concentric, past research has explored various specialized approaches targeting cost reductions in 3D printing. Sloped wall supports \cite{huang2009sloping}, tree-like space efficient approaches \cite{schmidt2014branching,meshmixer,vanek2014clever}, bridge supports \cite{shen2016bridge} and scaffolding style support structures \cite{dumas2014bridging} have been explored. While these work focus on reducing the build time and amount of material used to print supports, Curvy has an emphasis on enhancing the overall surface quality of the resulting print.  

In addition to aforementioned work on design of external support structures, variety of approaches has focused on developing internal supports for printing hollow objects \cite{hornus2016tight, hornus2017iterative, wu2016self, lee2017block}. While not our primary focus, our approach is inherently capable of generating internal support structures the same way it generates the external ones. 

\subsection{Build Orientation Optimization}
Effects of build orientation on build time and cost \cite{ahn2007fabrication,alexander1998part}, mechanical properties \cite{ulu2015enhancing}, surface roughness \cite{delfs2016optimized, wang2016improved}, manufacturability \cite{ulu2020manufacturability} as well as the support structure \cite{ezair2015orientation} and its impacts on the target object \cite{zhang2015perceptual} have been studied extensively and automated means to select the \textit{best} orientation are proposed to minimize such directional biases. Instead of selecting a single best orientation, other approaches use robotic printing platforms to manipulate the build orientation actively during the print process in order to avoid the need for support structures \cite{wu2017robofdm, wu2019general, dai2018support, gao2015revomaker, xu2019curved}. In our approach, we select a single build orientation that minimizes amount of support contact on surfaces that are intended to exhibit high surface quality when printed. 

Among all, Curvy is closest to \cite{zhang2015perceptual} that build orientation is adjusted to avoid support structures touching the perceptually important surfaces. However, our approach incorporates individual user's preferences rather than completely relying on a single generalized model. Curvy uses saliency map only as a starting point to guide the users in their selection. This approach allows Curvy to mitigate not only perceptual but also functional impact of supports on the target object.

%\begin{figure}
%\centering
%  \includegraphics[width=0.9\columnwidth]{figures/sigchi-logo}
%  \caption{Insert a caption below each figure. Do not alter the
%    Caption style.  One-line captions should be centered; multi-line
%    should be justified. }~\label{fig:figure1}
%\end{figure}

%\begin{table}
%  \centering
%  \begin{tabular}{l r r r}
%    % \toprule
%    & & \multicolumn{2}{c}{\small{\textbf{Test Conditions}}} \\
%    \cmidrule(r){3-4}
%    {\small\textit{Name}}
%    & {\small \textit{First}}
%      & {\small \textit{Second}}
%    & {\small \textit{Final}} \\
%    \midrule
%    Marsden & 223.0 & 44 & 432,321 \\
%    Nass & 22.2 & 16 & 234,333 \\
%    Borriello & 22.9 & 11 & 93,123 \\
%    Karat & 34.9 & 2200 & 103,322 \\
%    % \bottomrule
%  \end{tabular}
%  \caption{Table captions should be placed below the table. We
%    recommend table lines be 1 point, 25\% black. Minimize use of
%    table grid lines.}~\label{tab:table1}
%\end{table}
%-------------------------------------------------------------------------
\section{Curvy Support Design}

\subsection{Overview}
Figure~\ref{fig:fig2} illustrates the overview of our interactive support structure design process. Given an input 3D geometry in an arbitrary initial orientation, the user defines their preferences on desired surface quality by simply painting on the surface of the object. Then, Curvy optimizes the build orientation \one to avoid the support requirement at regions that are depicted to be high quality by the user as much as possible and \two to minimize the support contact area everywhere else. At the optimum orientation, the volume that is needed to be filled with support structure is computed and sliced into layers to generate \textit{support polygons} (green in Figure~\ref{fig:fig2}(b)). Note that layers here corresponds to layers/slices in 3D printing process. Then, each support polygon is filled with curvy zigzag toolpaths. Spacing between the individual toolpaths (\ie density) is adjusted based on user preferences on the surface quality. For higher quality surface regions, spacing is decreased in corresponding parts of support polygons below it. Smaller spacing between support toolpaths provide a higher quality bridging, thereby resulting in better surface quality. On the other hand, it increases the amount of material used for supports as well as the overall print time. To mitigate such effects in printing cost, density of supports are reduced in parts of support polygons corresponding to lower quality requirement areas. For example, note the density difference between the top left part of the slice and the rest of it in Figure~\ref{fig:fig2}(c). Finally, Curvy compiles a machine instruction (gcode) file by accumulating all the support toolpaths from each layer together with the toolpaths required to print the target object. To illustrate the resulting support structure as a whole, we construct a 3D model from the generated toolpaths in Figure~\ref{fig:fig2}(d).

\begin{figure*}
    \centering
    \includegraphics[width=\textwidth]{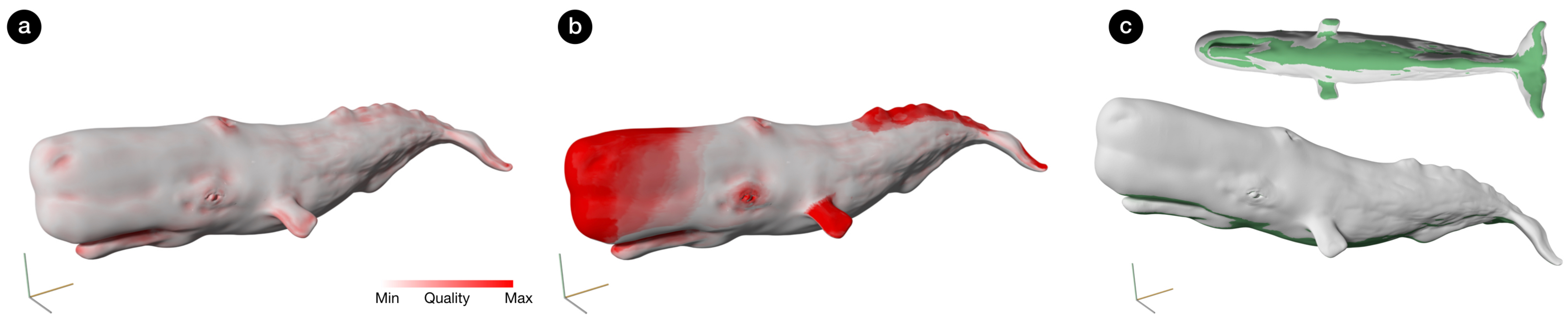}
    \caption{Overview of how user input is collected. As a starting point, mesh saliency is provided (a). User paints over it to convey their surface quality preferences (b). Curvy then optimizes the build orientation (c). Overhanging regions that require support are highlighted in green in bottom view inset.}
    \label{fig:ui}
\end{figure*}

Our motivation for generating curvy zigzags comes from two major observations in FFF: \one for high quality perimeter prints, a \textit{good} bridging distance should be maintained at the polygon boundaries and \two for easy removal, 
\begin{wrapfigure}{r}{0.3\columnwidth}
    \centering
    \includegraphics[width=0.3\columnwidth]{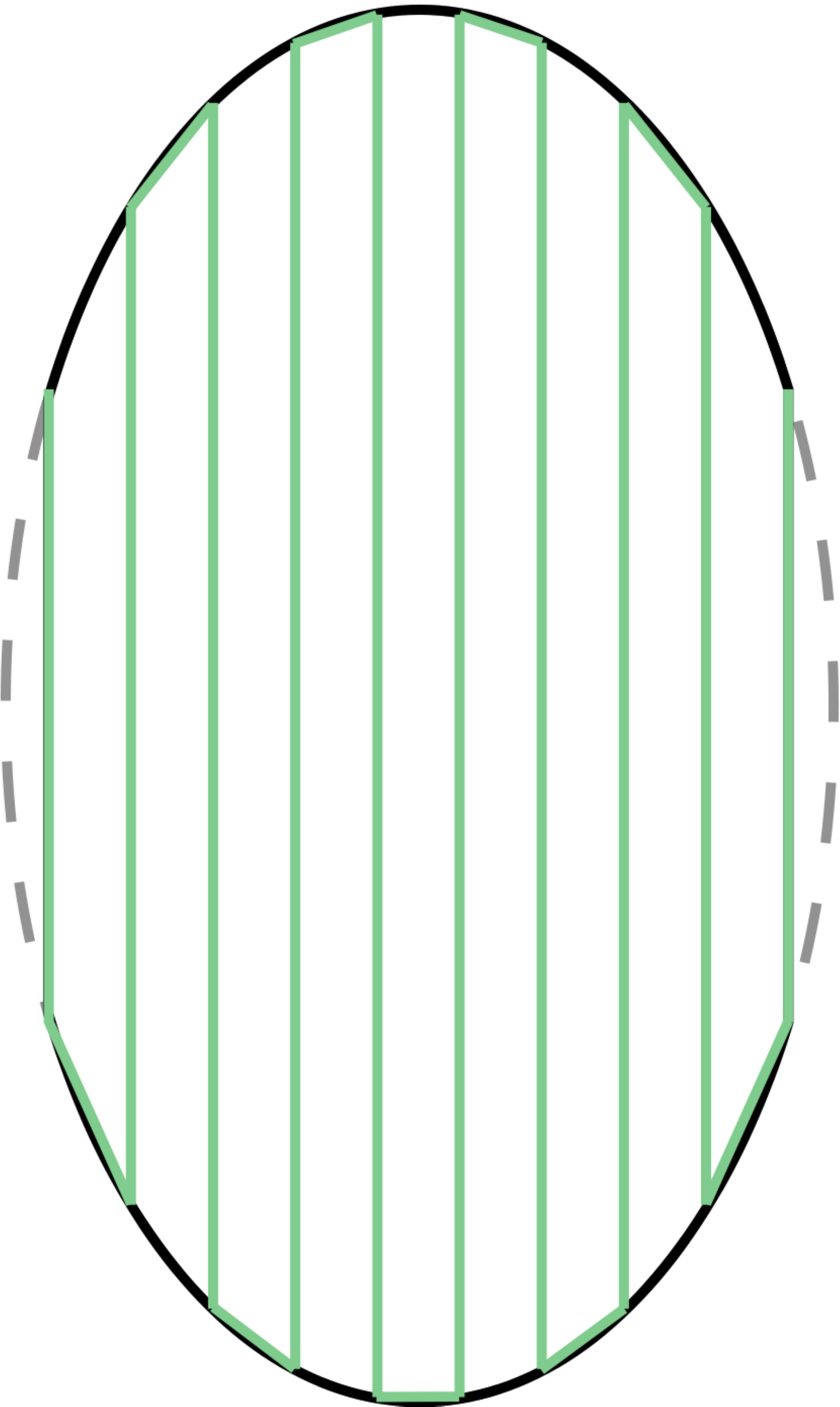}
%   \caption{Dashed parts of the perimeter are left unsupported for this particular zigzag orientation.}
\end{wrapfigure}
the support toolpaths should be perpendicular to both perimeter and infill toolpaths as much as possible. 
The former property comes from the fact that sagging will occur when the bridging distance between two supporting contact points is large. The latter one is due to the adhesion characteristics between the support and object layers. When the support paths are parallel to the infill or perimeter paths, the adhesion area increases, resulting in stronger bonding between them. 
The most commonly used regular zigzag support pattern does not satisfy these properties as the toolpaths are oriented in a predetermined direction. 
For an arbitrary shape boundary or arbitrary infill paths,
this direction may become parallel or close to parallel to them, leaving some parts unsupported (parts of the ellipse perimeter shown as dashed)
or resulting in very strong adhesion. Yet, both of these properties directly affect the surface quality as the perimeters mainly constitute the visible surface of the print and post-processing artifacts are often alleviated when the adhesion between the supports and the object is lower. Our curvy zigzag toolpaths demonstrate both of these properties by complying to a bidirectional field (Figure~\ref{fig:fig2}(c)) governed by the boundary of the polygon as well as the predetermined infill pattern.

\begin{figure}
\centering
  \includegraphics[width=\columnwidth]{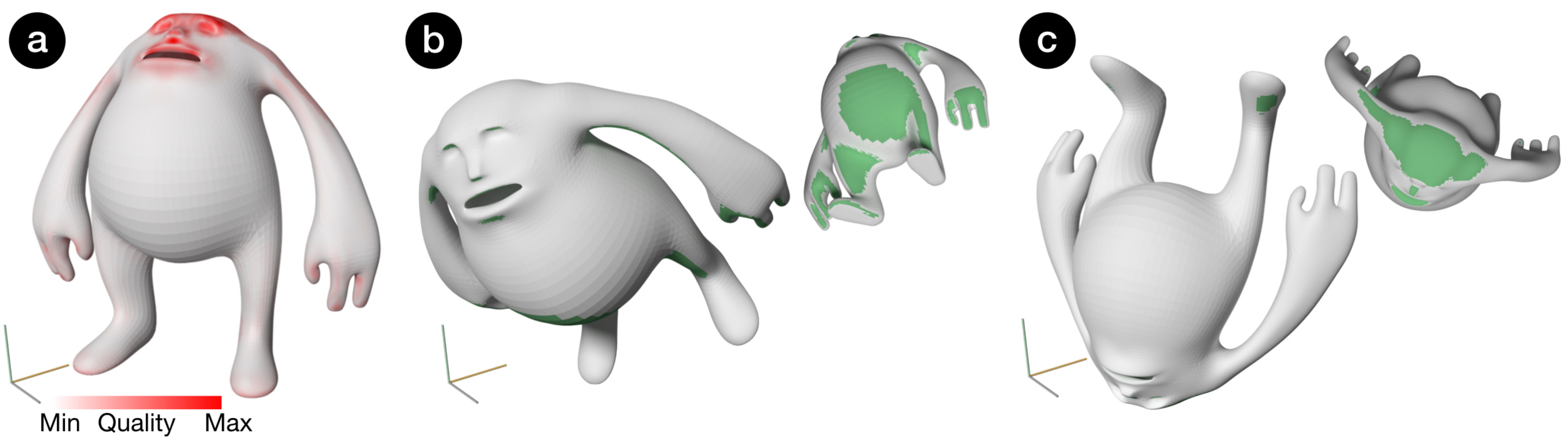}
  \caption{Effect of build orientation optimization weight. User preference (a) and corresponding optimum build orientation considering only user preference, $\omega = 1$ (b) and only support area, $\omega = 0$ (c). Overhang regions that require support are highlighted in green in bottom view insets.}~\label{fig:buildDirection}
\end{figure}

\subsection{Collecting User Preferences}

At the heart of Curvy lies the idea of implicit support design through collecting user preferences \textit{directly} on the target object rather than explicitly designing supports. Figure~\ref{fig:ui} provides an overview of user interaction and how user input is incorporated into support structure generation process. We start by computing mesh saliency on the input object such that salient features correspond to regions that require high surface quality Figure~\ref{fig:ui}(a). Then, the user paints over the saliency information expressing quality specifications for their own preferences and functional requirements Figure~\ref{fig:ui}(b). This quality information is, then, used to find the optimum build orientation such that support contacts are minimized avoiding high quality regions as much as possible. We store user preference as a scalar field $\boldsymbol{Q} \in [0,1]$ defined on the vertices of the object mesh representing \textit{quality} requirements. Then, $\boldsymbol{Q}$ is mapped to individual slices later during the support generation process where denser supports are created for regions with higher quality requirements.

We compute mesh saliency as presented in \cite{lee2005saliency}. The saliency computation utilizes curvature information on the input mesh and identifies visually interesting regions that are likely to be perceptually important. While perceptual saliency methods excel in detecting facial features such as eyes and nose, mesh saliency approaches often lack in capturing functionally important features that are crucial for 3D printed objects. For this reason, we allow users to paint over mesh saliency to express functional considerations. In that sense, saliency information maybe used as a starting point and guide users' preference. On the other hand, saliency information may partially or fully be removed on the regions that are not visually or functionally important to the user.

\subsection{Selecting the Build Orientation}

Contact surfaces where support structures touch the target object often have poor quality due to imperfect bridging or adhesion. Here, we present an optimization method to select a build orientation that results in minimal contact area at the regions designated to be high quality by the user.

\begin{figure*}
    \centering
    \includegraphics[width=\textwidth]{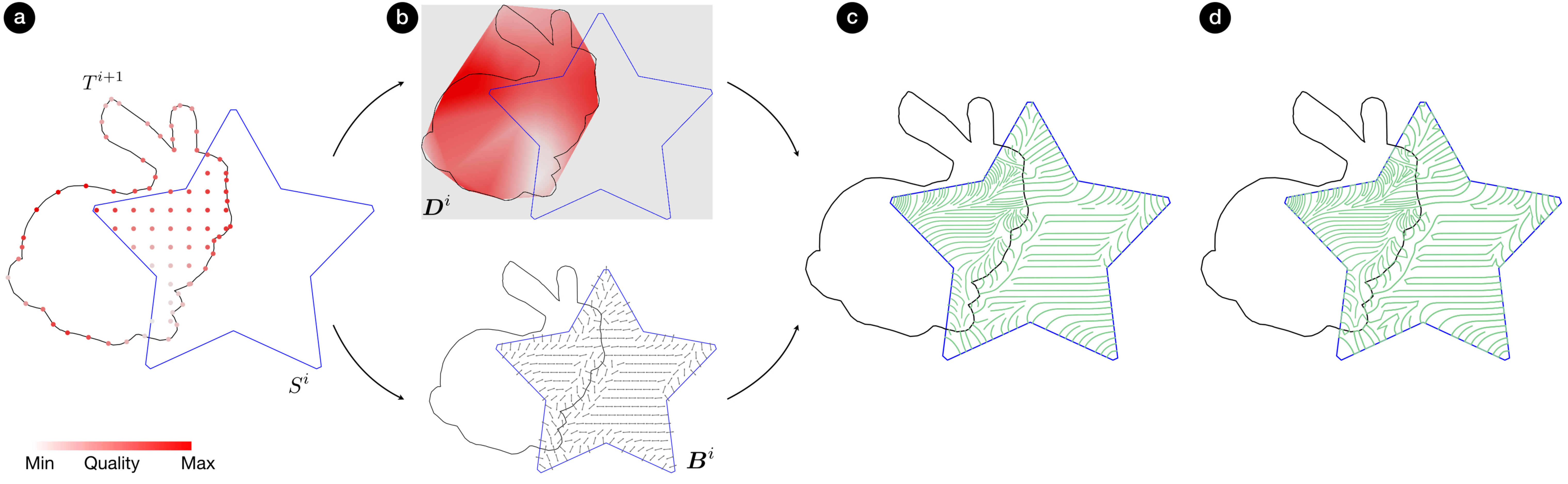}
    \caption{Given the support polygons and corresponding slice of the target object with the quality preferences (a), we construct a scalar density field (b-top) and a bidirectional field (b-bottom). We then integrate the bidirectional field to generate a set of streamlines. The spacing between the individual streamlines is dictated by the density field (c). Connected streamlines constitute the curvy zigzag toolpaths for this particular slice of the support structure (d).}
    \label{fig:fig5}
\end{figure*}

To determine if an overhang area requires support to be printed, how much the overhang tilts, $\varphi$ from the build direction,
\begin{wrapfigure}{r}{0.31\columnwidth}
\centering
    \includegraphics[width=0.3\columnwidth]{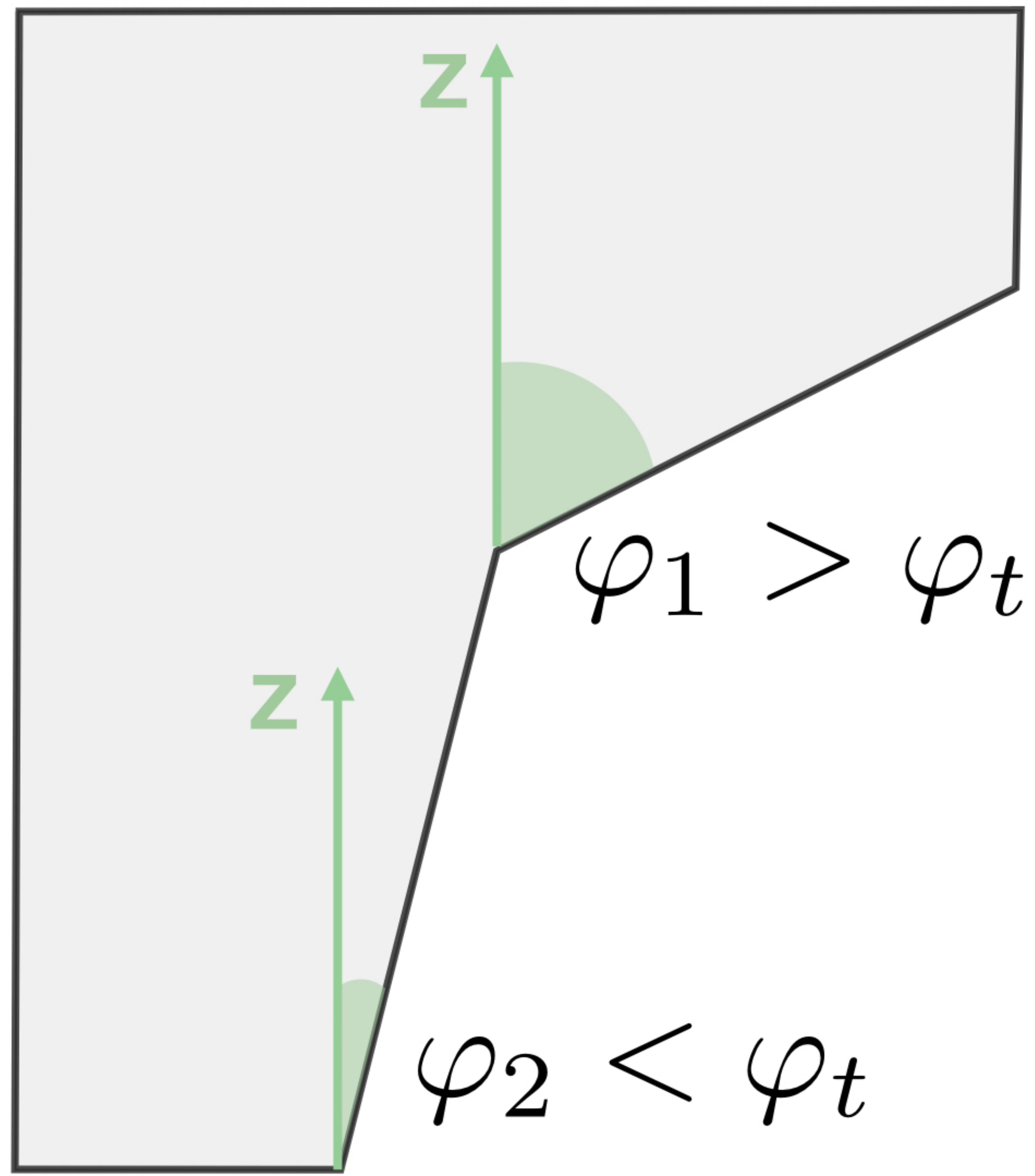}
  %\caption{Overhangs}
\end{wrapfigure}
$z$ is measured and compared to a threshold overhang angle, $\varphi_t$. If $\varphi>\varphi_t$, the overhang requires support structures. Using this principle, an object can be oriented to minimize overhang surface area that require support structures. Instead of simply minimizing the contact surface area, in this work, we incorporate user preferences to the objective function using the quality field, $\boldsymbol{Q}$. Since overhangs are defined on mesh faces, we map $\boldsymbol{Q}$ that is defined on the mesh vertices to mesh faces by averaging over vertices of each face, $q_j = \sum_k^{n_V} \boldsymbol{Q}_k / n_V$. Here, $q_j$ is the averaged quality value of face $j$, $n_V$ is the number vertices on the face, and $k$ is the vertex index. Then, we formulate our cost function as
\begin{equation}
\begin{aligned}
\kappa_j(\boldsymbol{\theta}) = 
    \begin{cases}
    0, & \text{if } \varphi(\theta) \leq \varphi_{t}\\
    (1-\omega)A_j + \omega A_j q_j^{1/p}, & \text{otherwise}.
    \end{cases}
\end{aligned}
\label{eq:objective}
\end{equation}
where $\boldsymbol{\theta} = [\alpha, \beta]$ represents rotations around z and x axes such that vertices of the rotated mesh can be calculated as $\boldsymbol{V}_r =\boldsymbol{R_x}(\beta)\boldsymbol{R_z}(\alpha)\boldsymbol{V}$. Here, $\boldsymbol{R_z}$ and $\boldsymbol{R_x}$ are rotation matrices, $\boldsymbol{V}$ is a matrix storing vertex positions, $A_j \in [0,1]$ is normalized area of face $j$, $p > 0$ is a penalization factor and $\omega \in [0,1]$ is a weight parameter. When $\omega$ is chosen to be 0, optimization minimizes total contact area without considering the quality input. On the other hand, when $\omega=1$, contact area overlapping with higher quality regions are minimized. In other words, optimization works towards eliminating any support contact at high quality regions. Figure~\ref{fig:buildDirection} demonstrates effect of $\omega$ parameter on an example case. While printing the object upside down (Figure~\ref{fig:buildDirection}(c)) results in the least amount of contact area (\ie $\omega=0$), this build orientation may not be desirable as the head is a perceptually important part of a figurine object.  

Formal definition of our build orientation optimization problem is as follows:
\begin{equation}
\begin{aligned}
& \underset{\boldsymbol{\theta}}{\text{min}} 
&& \sum_j^{n_F} \kappa_j(\boldsymbol{\theta}) \\
& \text{s.t.} && \alpha\in [-\pi,\pi] \\
             &&& \beta\in [0,\pi]\\
%& & & \boldsymbol{K(\boldsymbol{w})}\boldsymbol{u} = \boldsymbol{F}\\
\end{aligned}
\label{eq:buildOpt}
\end{equation}
where $n_F$ is the number of faces. In this formulation, only overhang angle, $\varphi$ needs to be recomputed for each objective evaluation as the object is reoriented. Thus, we have an optimization problem with a reasonably fast objective evaluation and only two optimization variables. We solve this problem using simulated annealing method~\cite{kirkpatrick1983sa} that finds global minimum when sufficient number of iterations are performed. For the examples of this paper, optimization converged to a solution in a few seconds achieving practical computation times for an interactive tool.

%-------------------------------------------------------------------------
\subsection{Generating Streamlines}

We construct curvy zigzag toolpaths as connected streamlines generated in a bidirectional field created inside a support polygon. Figure~\ref{fig:fig5} illustrates the main steps of our toolpath generation process. Let $S^i$ be a set of support polygons at layer $i$ and $T^{i+1}$ be the slice of the target object supported by $S^i$. We start by computing the quality requirements corresponding to the $i$th layer of our supports, $\bm{Q}^{i}$. This is done by simply taking a slab of $\bm{Q}$ between the layers $i$ and $i+1$, and projecting the per-vertex quality values on the surface of this slab onto the $i$th layer. This results in a set of samples on the perimeter and inside of $T^{i+1}$ with their corresponding scalar quality values (Figure~\ref{fig:fig5}(a)). Then, we construct two fields: \one a scalar density field $\bm{D}^i$ and \two a bidirectional field $\bm{B}^i$ (Figure~\ref{fig:fig5}(b)). The density field corresponds to the linear interpolation of the quality samples in $\bm{Q}^{i}$. The bidirectional field is obtained as the interpolation of the $T^{i+1}$ and $S^i$ boundary normals as well as a predetermined infill direction. Then, we generate streamlines to fill inside the support polygon by integrating $\bm{B}^i$ (Figure~\ref{fig:fig5}(c)).  Here, the spacing between the streamlines are adjusted locally according to the density field $\bm{D}^i$. Finally, the streamlines are trimmed to the boundaries of $S^i$ and neighboring ones are connected by their start or end points to create final curvy zigzag toolpaths (Figure~\ref{fig:fig5}(d)).

For a 3D model with $n$ slices (\ie $i \in [0, n]$ where $i=0$ and $i=n$ corresponds to bottom and top slices, respectively), we start from $(n-1)$th slice and process down to $i=0$. For any layer, if there exists a non-empty set of polygons $C^i = S^i \setminus T^{i+1}$, support toolpaths in this area is required to align with the support toolpaths in the layer above it, $S^{i+1}$ for a successful print. In order to guarantee such a property, we carry over the streamlines in $S^{i+1}$ to $S^i$ and trim them with $C^i$. Then, we only create new streamlines in the remaining region of $S^i$, $E^i$=$T^{i+1} \cap S^i$.

\subsubsection{Bidirectional Field}

Suppose the infill direction $\bm{f}$ is predetermined and let $\bm{d}$ be a vector perpendicular to it. In order to ensure that the streamlines generated inside $S^i$ are perpendicular to both $T^{i+1}$ perimeter and the infill inside it, we construct a field complying with the boundary normals of $E^i$ and $\bm{d}$. We first compute an \textit{effective infill region} by offsetting $E^i$ inwards. Inside this offset polygon set $O^i$, we enforce the field to be aligned with $\bm{d}$. On the boundary of $E^i$, we constrain the field to be aligned with the normals. Then, the region bounded by the boundaries of $E^i$ and $O^i$ constitutes the \textit{transition region} where the field orientation is obtained by linear interpolation. This allows us to obtain a smooth field inside $E^i$ to generate our streamlines. 

\begin{figure}
\centering
  \includegraphics[width=\columnwidth]{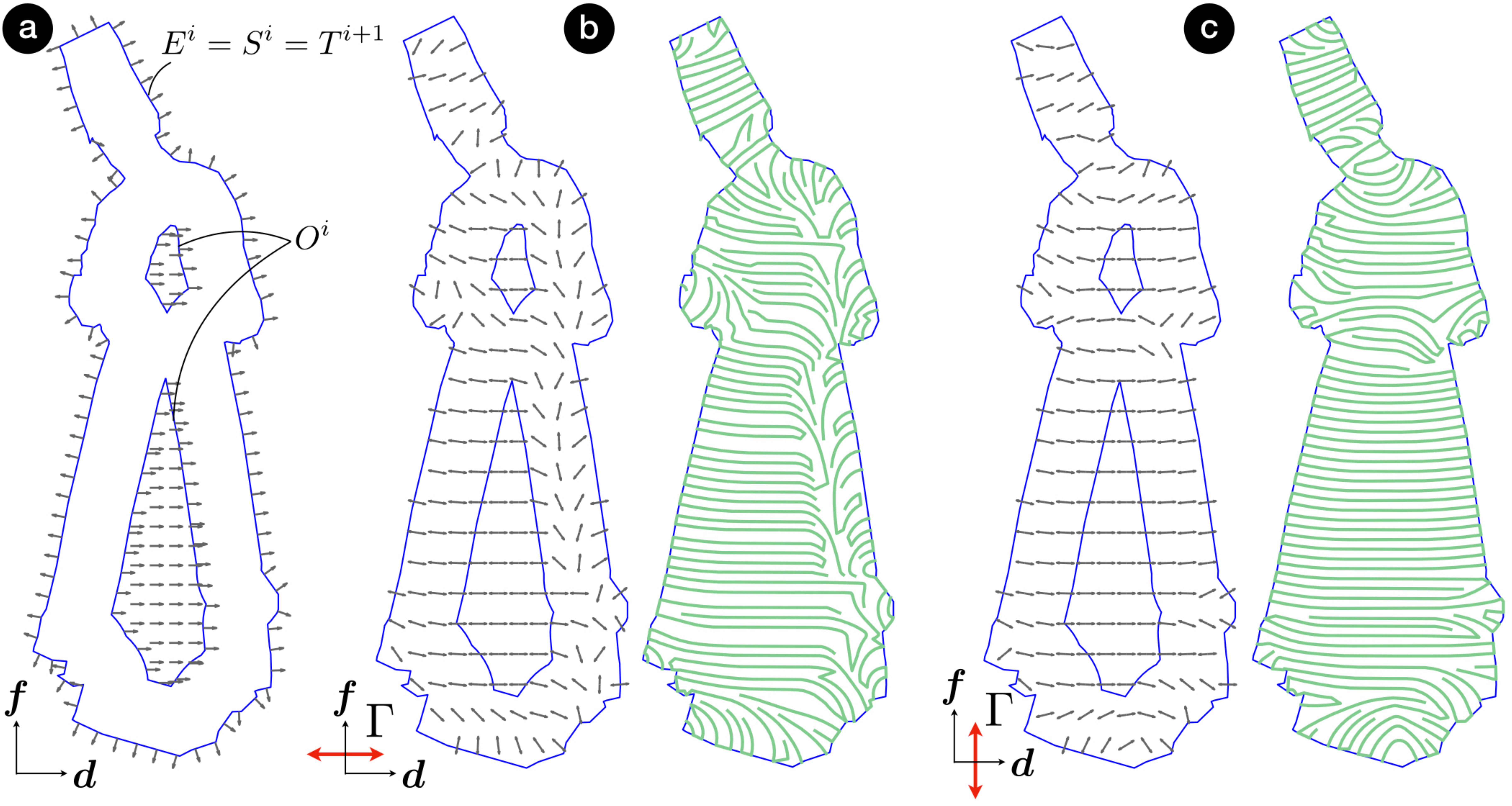}
  \caption{Construction of an example bidirectional field. For the same input support polygon (a), selection of $\Gamma$ along $\bm{f}$ results in longer and smoother paths in comparison to $\Gamma$ along $\bm{d}$ (b-c). }~\label{fig:fig6}
\end{figure}

Simply computing a vector field through interpolation of boundary normals and $\bm{d}$, however would result in a large number of singularity points that would prevent us from creating long and smooth paths. Consider a case where two points are located across each other on opposite sides of a rectangular polygon. As the normals of the polygon point outward, they are assigned vectors in opposite directions. Although there may be a single path connecting these two points while satisfying our criteria above, interpolation of a vector field in between these two points would result in a singularity point where the magnitude of the vector field becomes zero. Integration of such a vector field would result in a broken path between these two points. As the paths does not have directionality, we use a bidirectional field to avoid such problems. For bidirectional field interpolation, we represent each vector with the smallest angle it makes with a predetermined fixed axis, $\Gamma$. Figure~\ref{fig:fig6} demonstrates an example case. In this example, we assume $ E^i = S^i = T^{i+1}$ (\ie $C^i = \emptyset$ and there are no streamlines carried over from the $(i+1)$th layer) for simplicity. 

Bidirectional fields allow singularities in the neighborhood of which the field turns $\pi$ radians \cite{viertel2019approach}. In our formulation, this means that the singularities will occur around $\Gamma$. In order to minimize the number of singularities in the resulting field, we select $\Gamma$ along the infill direction $\bm{f}$. As the interpolation is often between the boundary normals of $E^i$ and $\bm{d}$, we observed that selecting $\Gamma$ this way keeps the active interpolation region away from the problematic area. An example comparison is provided in Figure~\ref{fig:fig6}(b) and (c). For the same input configuration, we obtain longer and smoother paths by selecting $\Gamma$ along $\bm{f}$ in comparison $\Gamma$ along $\bm{d}$. 

 In cases where $C^i \neq \emptyset$, we additionally use boundary normals of $S^i$ while computing the bidirectional field (see Figure~\ref{fig:fig5}). This allows us to expand the field to the entire $S^i$ and extend the streamlines towards $C^i$. We found such an approach particularly useful when $S$ (and therefore, $E$) is significantly small at a layer and gradually increases in size as moved towards the bottom. In such a case, streamlines in $E$ become too short in the top layers and get filtered out (as practically it may not be possible to print such short paths), thereby effectively skipping that particular support layer. In the following layers below it, extending streamlines to entire $S$ allow us to compensate for the filtered out streamlines and prevent skipping layers that have large $S$ to generate long enough streamlines.

\subsubsection{Streamline Spacing}
We adopt a similar approach to \cite{jobard1997creating} in creating streamlines with controlled density. Different from this method, our approach controls the spacing between adjacent streamlines locally using the density field $\bm{D^i}$ rather than a global density parameter. We generate streamlines by performing numerical integration of the bidirectional field $\bm{B}^i$. Seed point for a new streamline is chosen at a distance $d_{sep}$ away from an existing streamline. As the streamlines are represented as polylines (series of points), this corresponds to simply offsetting a point on a streamline in its normal direction. Then, starting from the seed point, a new streamline is iteratively elongated in both directions until it hits the domain boundary (\ie boundary of $S^i$), reaches to a singular point where magnitude of the field is close to $0$ or comes closer to another existing streamline than a distance $d_{test}$. Streamline generation stops when there is no more valid seed points. In our approach, an arbitrary point on the boundary of $E^i$ is selected as the starting seed point to initialize the algorithm. In order to avoid leaving large gaps or skipping unconnected components of $S^i$, we use additional seed points placed on a regular grid created inside the bounding box of $S^i$. 

\begin{figure}
\centering
  \includegraphics[width=\columnwidth]{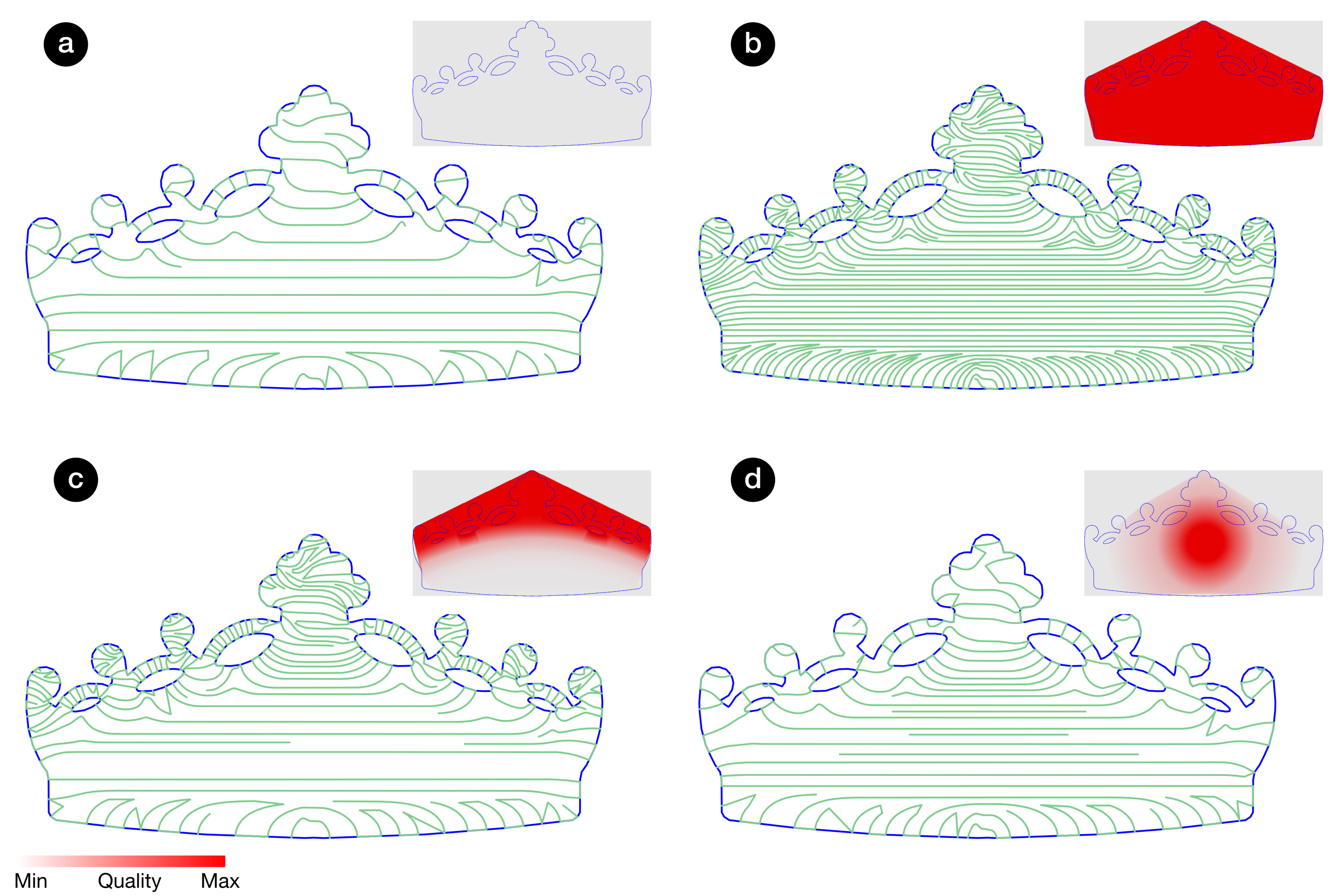}
  \caption{Effect of density fields on generated toolpaths: constant density fields (a)-(b) versus arbitrary varying density fields (c)-(d).}~\label{fig:fig7}
\end{figure}

In order to control the local spacing between the streamlines, we compute $d_{sep}$ as
\begin{equation}
    d_{sep} = d_{min} + (1 - \bm{D^i}(x,y))  (d_{max} - d_{min})
\end{equation}
where $\bm{D^i}(x,y)) \in [0,1]$ is value of the density field at a location $(x,y)$. Here, $d_{min}$ and $d_{max}$ are minimum and maximum allowed spacing, respectively. $d_{max}$ is dictated by the process and corresponds to the maximum distance for successful bridging. On the other hand, $d_{min}$ may theoretically be $0$, corresponding to solid filling of the support area. In our examples, we set $d_{min}=0.3 d_{max}$ to avoid impractically strong adhesion between the supports and the object. We use $d_{test}$ as a percentage of $d_{sep}$. This relaxes the constraints on streamline generation and helps us obtain longer streamlines by increasing the minimal distance at which the integration of the streamline will be stopped \cite{jobard1997creating}. We found that $d_{test} = 0.7 d_{sep}$ provides us a good balance between obtaining longer streamlines and having a stricter control on the spacing between streamlines. Figure~\ref{fig:fig7} illustrates the effect of density field on the spacing between streamlines. Toolpaths generated for arbitrary $\bm{D^i}$'s are demonstrated (Figure~\ref{fig:fig7}~(c)-(d)). Two extreme cases where $\bm{D^i} = 0~\forall (x,y) \in S^i$ (\ie $d_{sep} = d_{max}$) and $\bm{D^i} = 1~\forall (x,y) \in S^i$ (\ie $d_{sep} = d_{min}$) are also provided as baseline cases (Figure~\ref{fig:fig7}~(a)-(b)). 

\begin{figure}
\centering
  \includegraphics[width=\columnwidth]{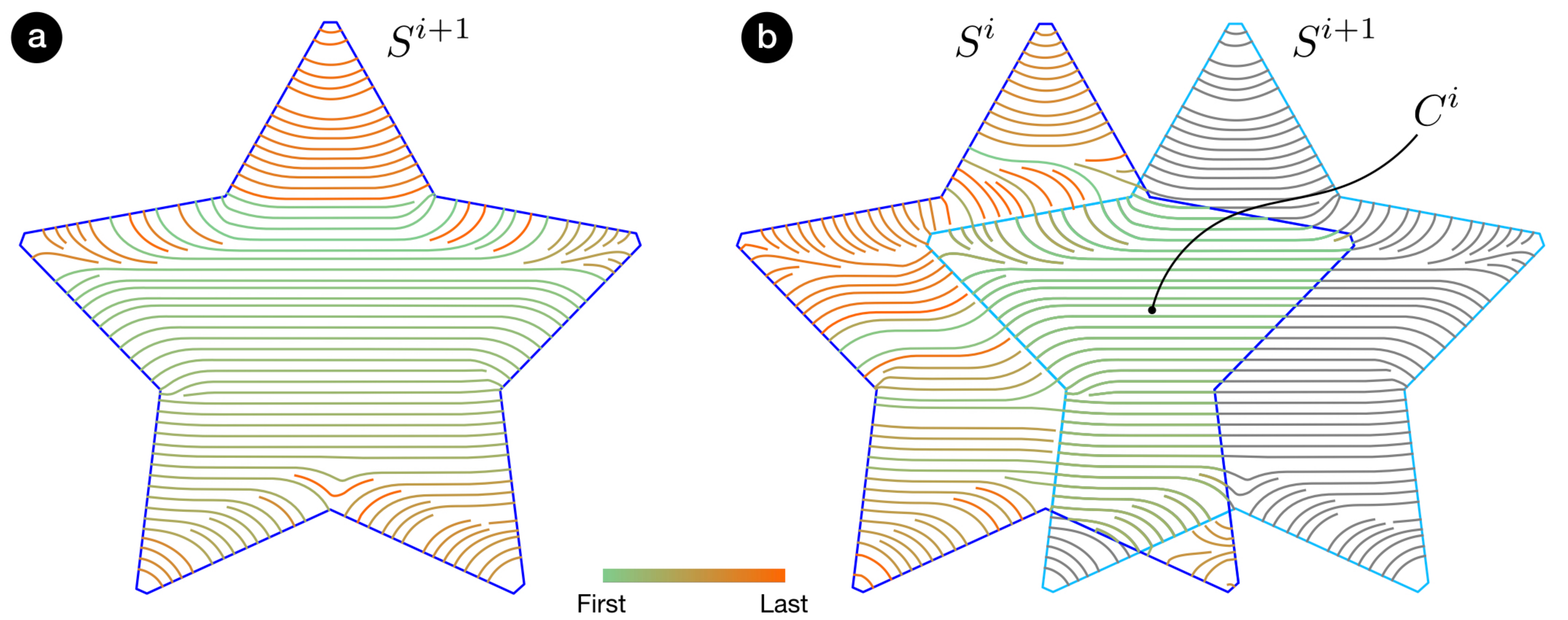}
  \caption{The order at which each streamline is created reveals our seed point selection approach. Different from  $C^{i+1} = \emptyset$ (a), end points of carryover streamlines are used as seed points first when $C^i \neq \emptyset$ (b). Gray parts of the streamlines belong to $S^{i+1}$ and are not carried over to $S^{i}$.}~\label{fig:fig8}
\end{figure}

For the integration, we use a fixed step Euler integrator where each new point of a streamline is calculated as
\begin{equation}
    p_{k+1} = p_k + h \bm{\bar{b}}(p_k).
\end{equation}
Here, h is the integration step and $p_k$ is the last point of the streamline. $\bm{\bar{b}}(p_k)$ represents the unit direction vector at point $p_k$ and it is obtained from the field $\bm{B}^i$. As $\bm{B}^i$ is bidirectional, there are two possible directions to extend the streamline at any arbitrary point $p_k$. Suppose $\bm{B}^i(p_k) = \{+\bm{b}, -\bm{b}\}$ where $\bm{b}$ is a vector. Among the two possible direction, we select the one that deviates the least from the previous step as
\begin{equation}
    \bm{\bar{b}}(p_k) = \argmax_{\bm{x} \in \bm{B}^i(p_k)} (p_k - p_{k-1}) \cdot \bm{x}. 
\end{equation}
In order to avoid streamline making abrupt turns, we stop the integration when the angle between $\bm{\bar{b}}$ and the last line segment of the streamline is larger than a certain threshold. In our examples, we found $\pi/3$ radians to work well as this threshold value. 

When $C^i \neq \emptyset$, we first elongate the streamlines that are carried over from $S^{i+1}$ as much as possible before starting to generate new streamlines in $E^i$. For this purpose, end points of carryover streamlines are used as seed points for the integration. New seed points are then created by offsetting the elongated streamlines. Figure~\ref{fig:fig8} demonstrates an example case. Order at which each streamline is created reveals our seed point selection process.

\subsubsection{Toolpath Generation}
Given the streamlines in each layer of support $S^i$, the last step in our support generation approach involves connecting neighboring ones to generate long and continuous curvy zigzag paths. To do that, we first trim streamlines to the boundary of $S^i$. Then, starting from the shortest streamline, our algorithm visits each streamline $l_j$ in order and connects it to another unvisited streamline $l_k$ in close proximity. For a successful connection to happen between $l_j$ and $l_k$, we look for the following criteria: \one an end point $l_j$ is sufficiently close to an end point of $l_k$, \two the connecting path does not intersect with other streamlines or other connecting paths and \three the connecting path does not leave the boundaries of $S_i$. When a successful connection occurs between $l_j$ and $l_k$, extension continues from the opposite end of $l_k$ until there is no more valid connection available.

\begin{figure}
\centering
  \includegraphics[width=0.9\columnwidth]{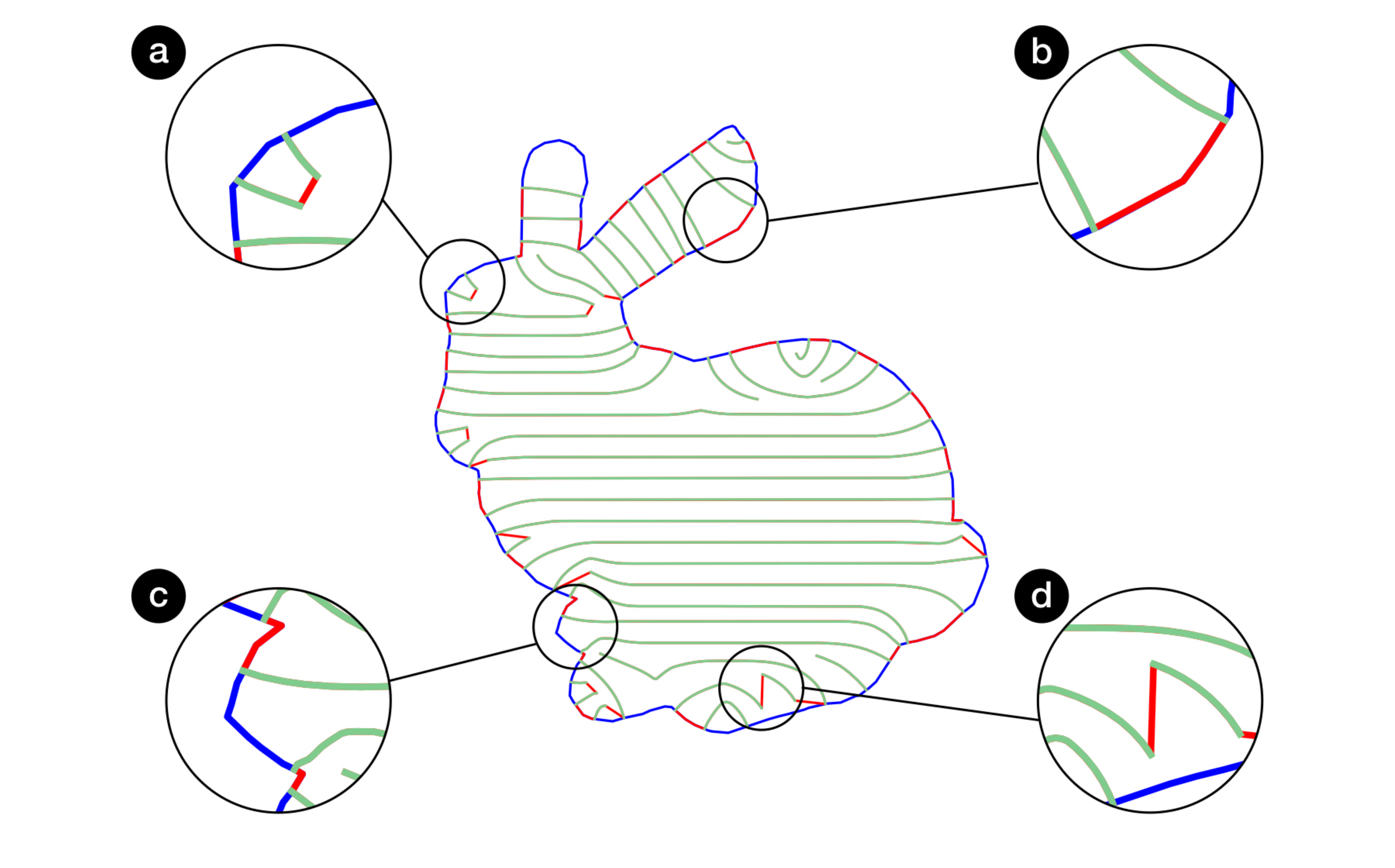}
  \caption{Two types of connecting paths: Straight path (a)-(d) and boundary following path (b)-(c). All connecting paths are shown in red.}~\label{fig:fig9}
\end{figure}

In our algorithm, we create two types of connecting paths between streamlines-- straight path and boundary following path. The former one is created when the connecting end points of both $l_j$ and $l_k$ are inside $S^i$. In this case, the end points are simply connected with a straight line. The latter one is created when the connecting end points are on the boundary of $S^i$. This time our algorithm uses the shortest boundary segment connecting these two points as the connecting path. This approach allow us to create toolpaths that supports the perimeters well and maintains the boundary details. An example case is illustrated in Figure~\ref{fig:fig9}.

For slices with $C^i \neq \emptyset$, we do not carryover the connecting paths and re-evaluate the connections after all the streamlines are created in the current slice. This is mainly because \textit{better} connections may be created after the carryover streamlines are extended in the new layer. Here, the term better often indicates shorter connection distance.

\begin{figure*}[!h]
    \centering
    \includegraphics[width=\textwidth]{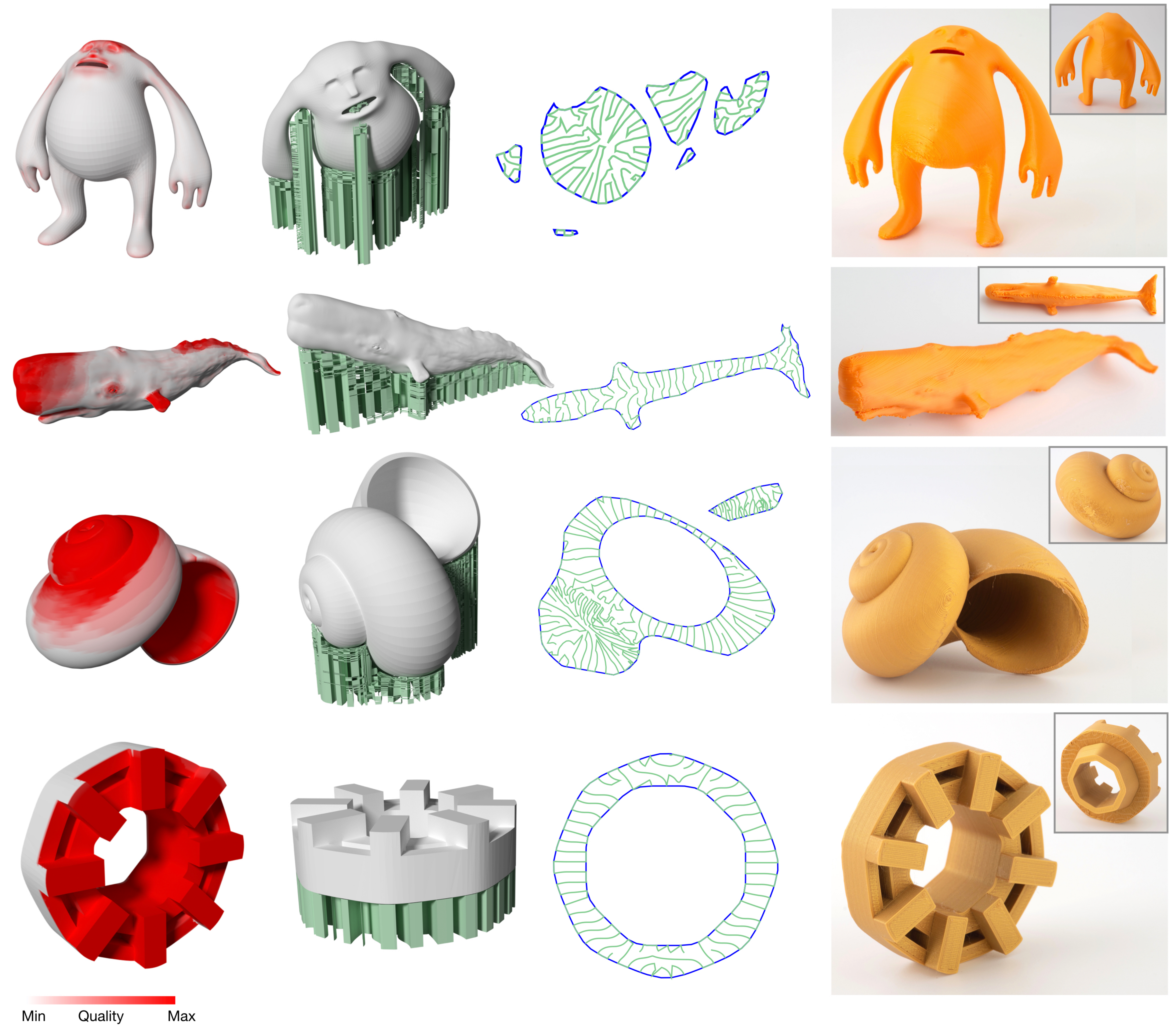}
    \caption{Example results. Left-to-right: quality preferences, resulting supports in optimum orientation, example slices of support structures and 3D printed results.}
    \label{fig:results}
\end{figure*}

In the presence of carryover streamlines, C\textsuperscript{1} continuity is often broken, reducing to C\textsuperscript{0}, during the elongation process described earlier. The reason behind this is that the bidirectional fields often change from one slice to another. For example, carryover streamlines are created by integrating $\bm{B}^{i+1}$ while the extensions are done in a new field $\bm{B}^i$. In such cases, we smooth the streamline around the junction point using Laplacian smoothing \cite{botsch2010polygon}. Resulting paths are more suitable for FFF type printing as issues related to acceleration/deceleration at sharp corners are eliminated, resulting in an improvement in print time. 

After the individual paths are created, machine instructions are compiled in the form of commonly accepted gcode. Here, both our support toolpaths and the target object toolpaths are accumulated together for all layers. We use gsSlicer\cite{gsSlicer} to generate toolpaths for the target object as well as to convert all the toolpaths to gcode.

%------------------------------------------------------------------------
\section{Results and Discussion}

\begin{figure}
\centering
  \includegraphics[width=\columnwidth]{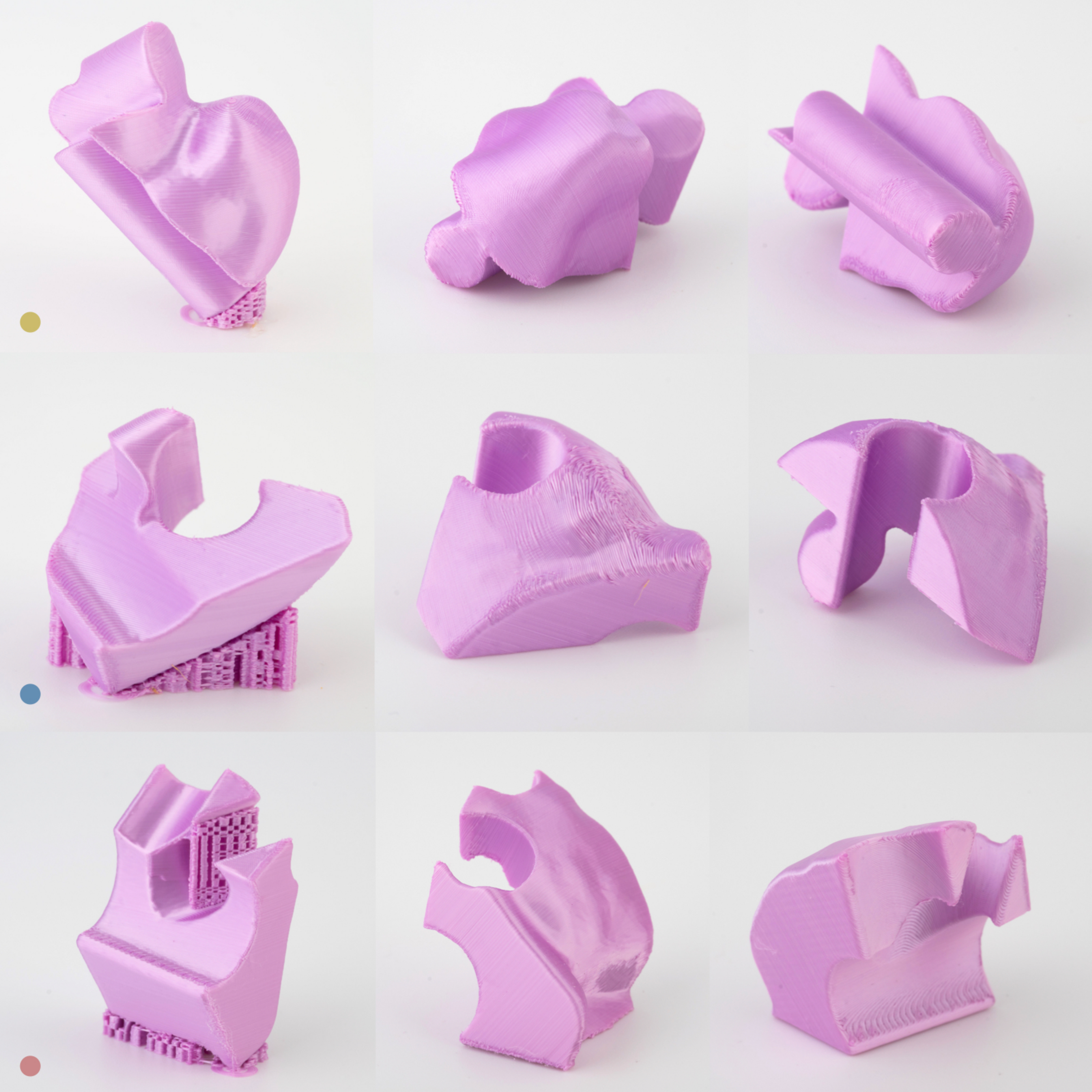}
  \caption{Detail views of 3D printed bunny puzzle pieces. Each row is marked with the color indicating correspondence to Figure~\ref{fig:teaser}. }~\label{fig:bunnyDetail}
\end{figure}

\begin{figure}[h]
\centering
  \includegraphics[width=\columnwidth]{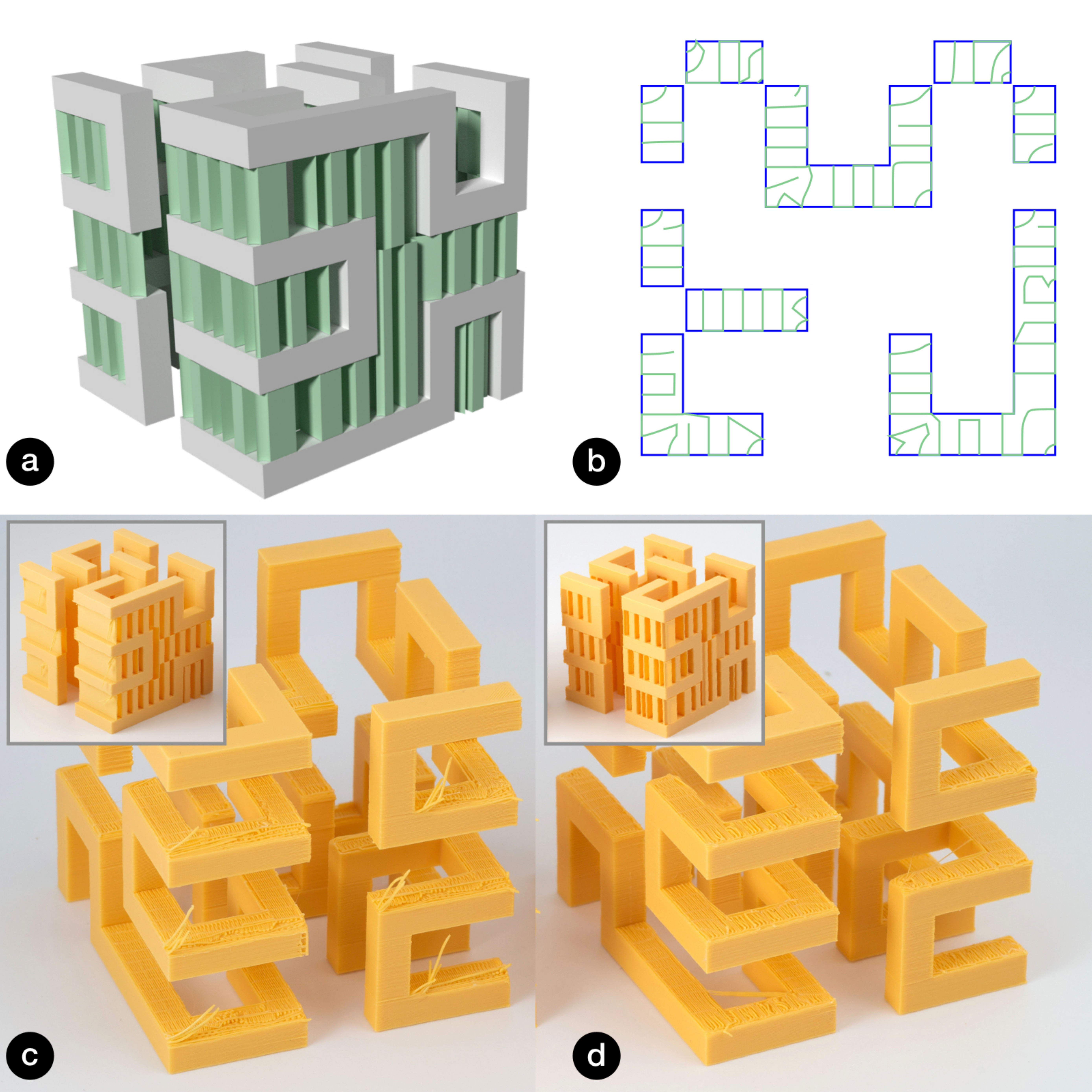}
  \caption{Hilbert Cube. Curvy zigzag supports (a), an example slice of the support structure (b). Comparison of 3D printed results using regular zigzag (c) and our curvy zigzag (d) supports. Inset figures in (c)-(d) shows the printed models before support removal.}~\label{fig:hilbertComparison}
\end{figure}

\begin{figure}[!h]
\centering
  \includegraphics[width=\columnwidth]{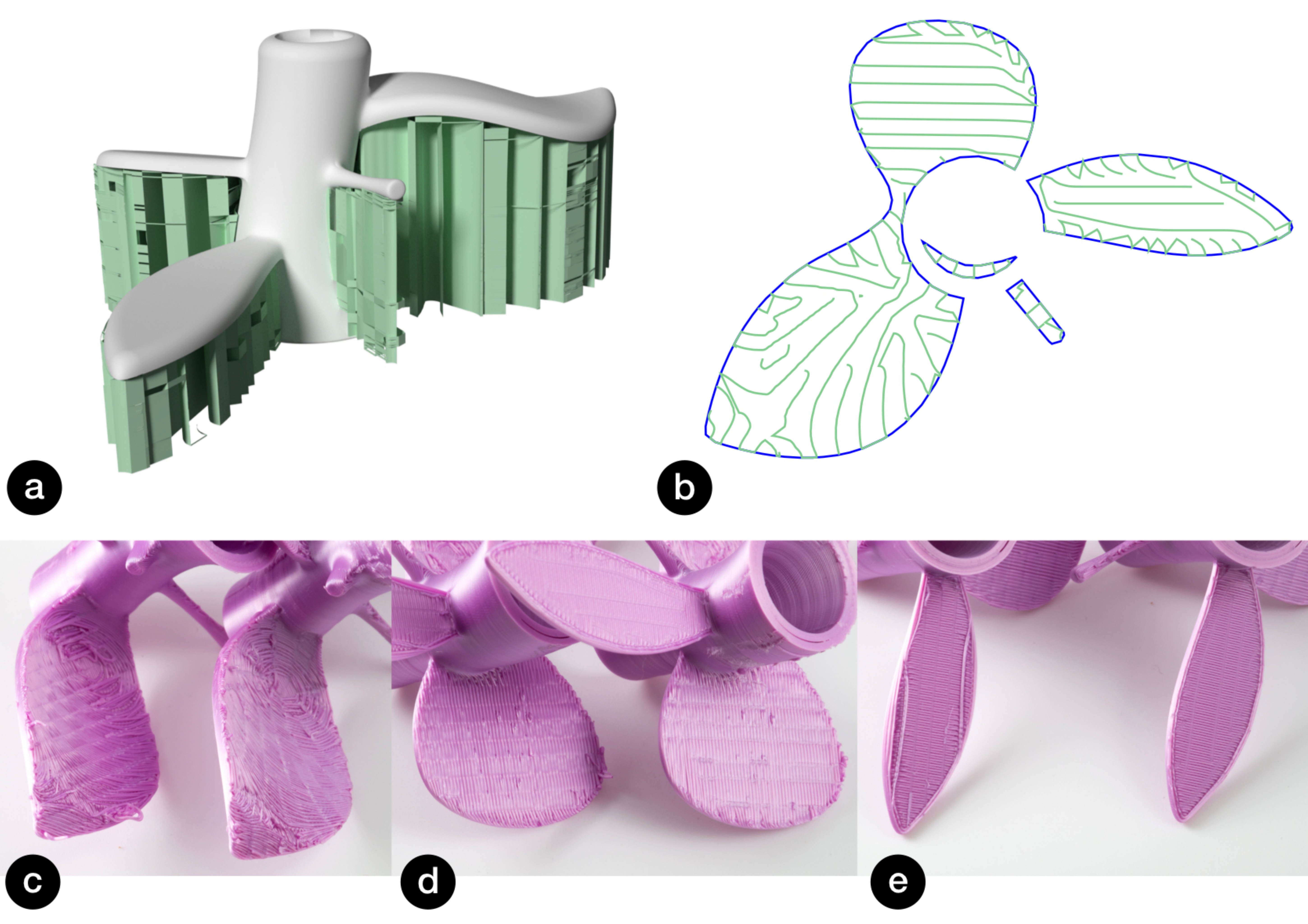}
  \caption{Plant. Curvy zigzag supports (a), an example slice of the support structure (b). Comparison of 3D printed results using regular zigzag (left) and our curvy zigzag supports (right) in (c)-(e).}~\label{fig:plantComparison}
\end{figure}

\begin{figure}[!h]
\centering
  \includegraphics[width=\columnwidth]{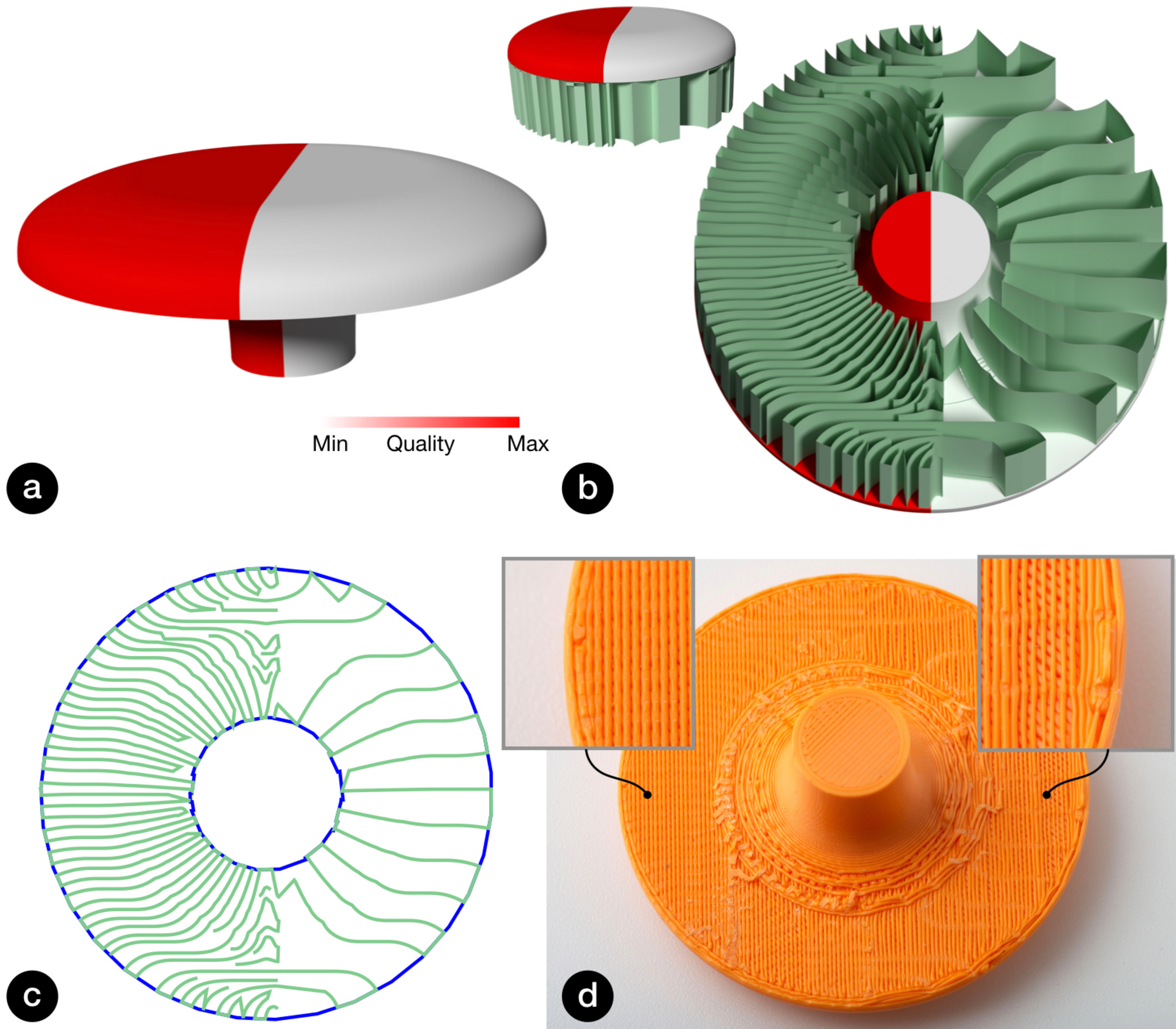}
  \caption{Mushroom. Desired surface quality (a), resulting curvy zigzag supports (b), an example slice of the support structure (c) and 3D printed result demonstrating the effect of the input user preferences on the surface quality (d). Close-up views of the surface are given in insets in (d). }~\label{fig:effectOfDensity}
\end{figure}

We demonstrate the performance of our method on a variety of models and validate it with 3D printed results. For all the examples, we set $\bm{f}=[0,1]$, $\varphi_t = \pi/4$, layer height to be $0.125mm$ and $\omega=1$, unless otherwise stated. Resulting structures are printed on a custom reprap FFF printer using PLA material. In order to show the resulting surface quality objectively, we remove the supports roughly without performing any detailed cleaning or post-processing.

Figure~\ref{fig:results} illustrate example results obtained using Curvy. All the example models are successfully printed in corresponding optimal orientations with curvy zigzag supports. For cases where the build orientation optimization is able to avoid supports in all high quality regions (\eg~ Mr. Humpty model in the first row, or mechanical model in the last row), curvy zigzags are generated with constant spacing of $d_{max}$. On the other hand, when there is no possible orientation that supports can be avoided at high quality regions all together, significant density variations are observed. For example, notice the dense region on the bottom left area of the example slice in the shell model (third row) in comparison to rest of the slice. 

In our approach, user selection may be driven by visual preferences (\eg first three rows of Figure~\ref{fig:results}) as well as functional requirements. Figure~\ref{fig:teaser} and last row of Figure~\ref{fig:results} demonstrate example cases where the motivation for the user selection is mainly functional. In these examples, the surfaces that are required to have tight geometric tolerances due to contact with other parts are assigned high quality values. In resulting prints, supports contacting those surfaces are completely avoided or designed to have high density curvy zigzags to achieve best print quality. Figure~\ref{fig:bunnyDetail} shows detail views of 3D printed puzzle pieces. For the face piece (marked with yellow), perceptually important face as well as functionally important tabs and grooves are marked as high quality regions. Therefore, they are left support free as much as possible. For other pieces (marked with blue and red), only tabs and grooves are assigned high quality requirements, therefore some of their outward facing surfaces show support contact marks in resulting prints. 

\subsection{Comparison}
We compare the performance of our curvy zigzag supports with the regular zigzags generated using Simplify3D\cite{simplify3d}. For an accurate comparison, we do not optimize the build orientation and we use a constant $d_{sep}$ that is equivalent to regular zigzag spacing. In Figure~\ref{fig:hilbertComparison} and Figure~\ref{fig:plantComparison}, we illustrate example cases. In Hilbert cube example, our curvy zigzags provide better support to overhang surfaces by changing the zigzag orientation to comply with their boundaries (Figure~\ref{fig:hilbertComparison}(b)) as opposed to constant orientation in regular zigzags. As a result, bridging distance is well maintained overall, thereby allowing a better quality print. Similarly, in plant model, our curvy toolpaths result in significantly better perimeters while quality degrades on perimeter regions that are close to parallel to zigzag direction in regular zigzags. 

For the same spacing value, we measured $2\%$ increase in total print time and $10\%$ increase in total material amount with curvy zigzags in comparison to regular zigzags. This increase mainly comes from the fact that curvy toolpaths can cover an arbitrary polygon better and leave smaller amount of empty regions compared to regular zigzags aligned on a predetermined grid.  

\subsection{Density}
We demonstrate the effect of support density on the surface quality on an example model in Figure~\ref{fig:effectOfDensity}. We set half of the mushroom model to have the highest quality while the other half is assigned the lowest quality value. We observe significant improvement on the perimeters as well as the infill region on the half supported by dense curvy zigzags over the other half supported by low density zigzags. On the low density half, gaps are formed between individual extrusions due to sagging in the infill region and accuracy diminishes due to large bridging distance on the perimeter. Yet, low density half of the supports can be printed using $\sim70\%$ less material which translates to similar savings in build time. 

%In our approach, build orientation optimization allow users to benefit from such possible savings. It favors the use of low density supports over high density ones by automatically selecting the orientation that would require least amount of supports on perceptually or functionally important regions of the target object. Therefore, supports are shifted towards regions with low quality requirements, often resulting in less material usage and shorter build time. Figure~\ref{fig:effectOfOrientation} illustrates an example case. In this particular model, we achieved \erva{$X\%$} reduction in material use and \erva{$X\%$} improvement in built time by optimizing the orientation to minimize the use of supports in high quality requirement regions.

\subsection{Design Tool}
We implemented the user interface in C++ and the backend toolpathing algorithm in Python. We use Shapely\footnote{https://pypi.org/project/Shapely/} for geometric operations and SciPy\footnote{https://www.scipy.org/} for field interpolations. 

\begin{figure}
\centering
  \includegraphics[width=\columnwidth]{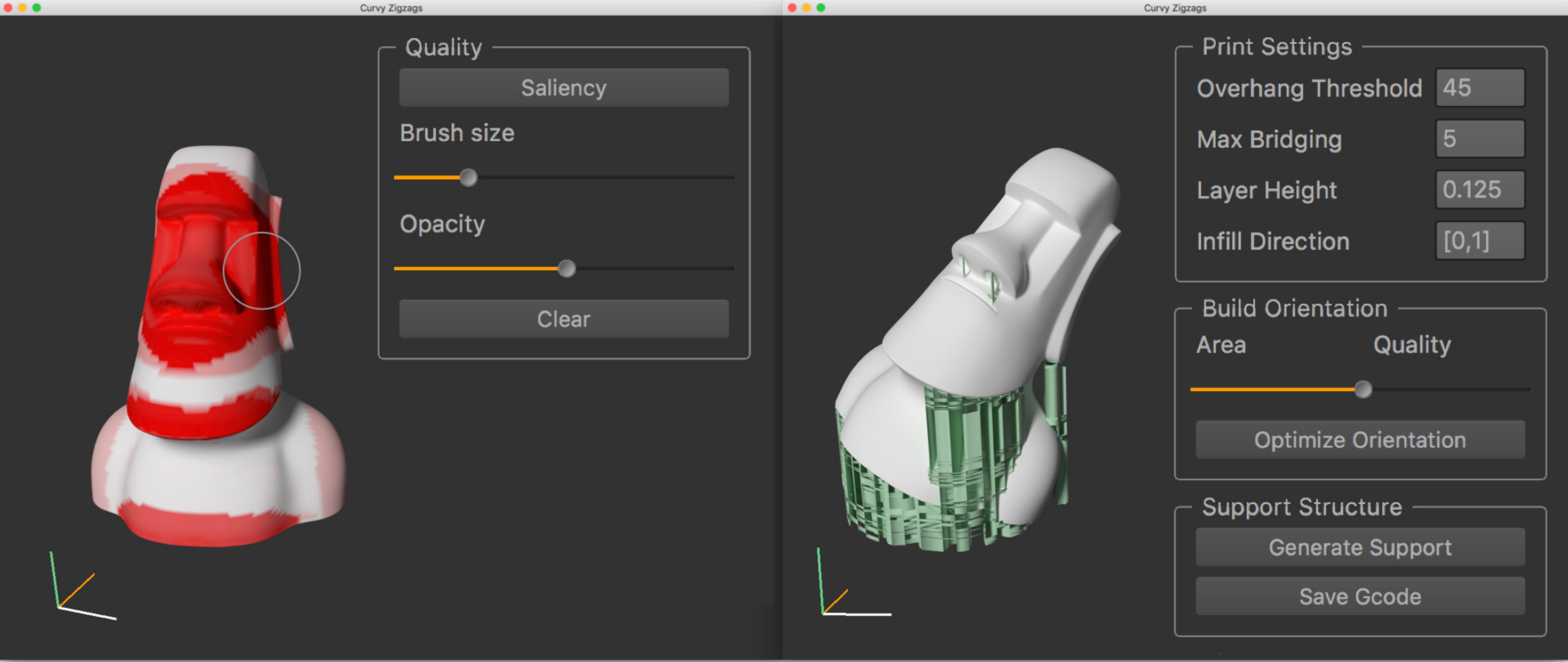}
  \caption{Interface of our design tool. Quality toolkit for user input (left) and build orientation optimization and support structure generation options (right).}~\label{fig:designTool}
\end{figure}

Figure~\ref{fig:designTool} demonstrates the interface of our design tool. After providing the surface quality input, users have an option to perform the build orientation optimization. For the support toolpath generation, users can calibrate the system to their own printer by adjusting $d_{max}$, $\bm{f}$, $\varphi_t$ and layer height. The system outputs G-code files.

\subsubsection{Preliminary User Feedback}
In order to collect user feedback to evaluate and validate Curvy, we conducted an informal, guided preliminary user study with 8 participants (aged 16-48). 7 out of 8 participants have reported that they are familiar with 3D printing and 6 of them have used a 3D printing software that is capable of generating support structures of some form.

\begin{figure}
\centering
  \includegraphics[width=\columnwidth]{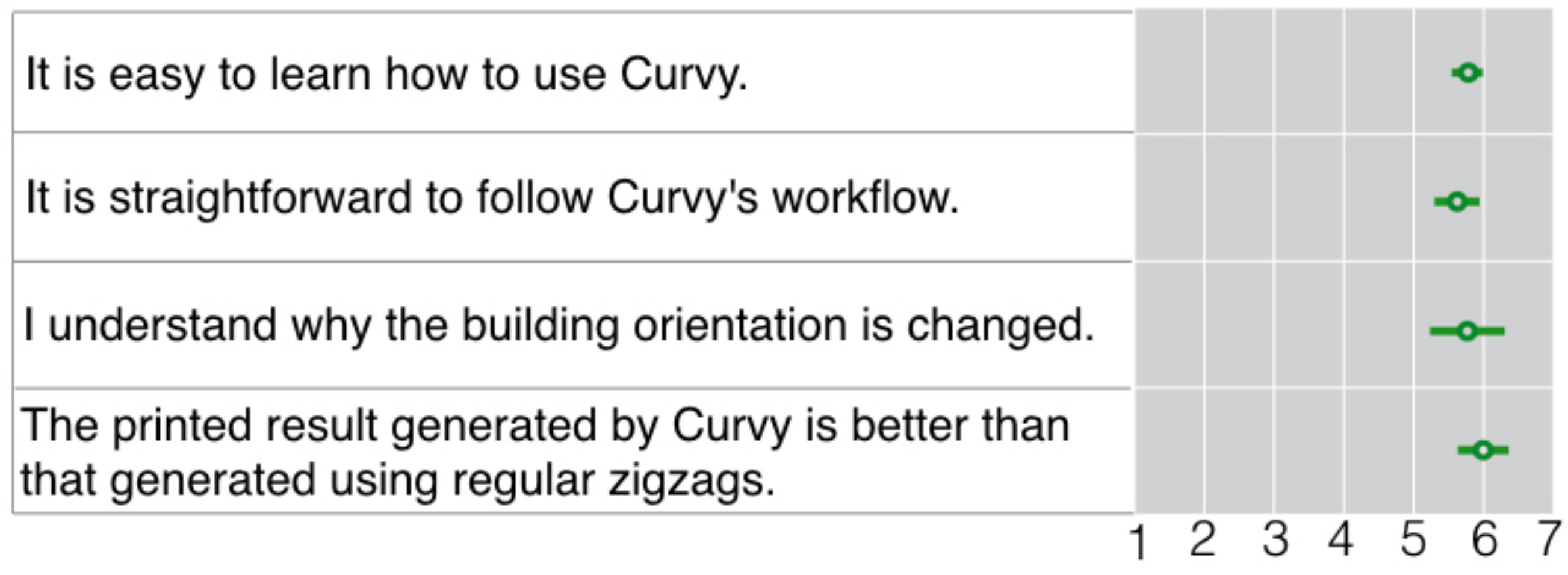}
  \caption{Survey results for selected questions (1: Strongly Disagree, 7: Strongly Agree). Circles and error bars represent mean values and standard errors, respectively.}~\label{fig:userStudy}
\end{figure}

The participants joined the 45 minute remote study where they were introduced the concept of support structures in 3D printing and their effects on surface quality. Then, they are shown a video demonstrating support structure design process in Simplify3D\cite{simplify3d} as well as Curvy on the Moai Statue model. Once they are familiar with the tools, they are asked to use Curvy to design support structures on an arbitrary model. Participants are asked to use their favorite 3D model or the provided Moai Statue model. They were also shown 3D printed results of Hilbert Cube (Figure~\ref{fig:hilbertComparison}), plant (Figure~\ref{fig:plantComparison}) and mushroom (Figure~\ref{fig:effectOfDensity}) models to compare the resulting surface quality with regular zigzag supports. Finally, participants are asked to complete a survey evaluating their overall experience with Curvy and quality of the resulting prints by comparing them with ones obtained using regular zigzag supports (in the Likert scale 1-7). 

Survey results for a selected set of questions are reported in Figure~\ref{fig:userStudy}. Overall, we found that the participants saw a clear benefit of using Curvy for designing support structures. They reported that Curvy is easy and straightforward to use. Yet, for some, adjustment of build orientation optimization weight was not intuitive. Participants found the painting style input easy and intuitive to express quality expectations. They reported to like the immediate feedback they are getting as opposed to conventional way of explicit support design. One important comment we received was that such free-form surface painting was useful for organic shapes but for man-made geometries such as mechanical models, one participant suggested that a surface selection tool would be useful for better precision. In general, participants expressed that the print quality obtained using Curvy is better than what is achieved with regular zigzags. $50\%$ of them indicated strong preference of using Curvy over other conventional software for designing support structures. Only one of the participants reported to prefer conventional software as opposed to Curvy.

\subsection{Limitations and Future Work}
Our curvy toolpath generation algorithm is most suitable for sparse filling of support polygons. For dense filling, it may create large number of short streamlines. Although this may be technically viable, it may not be practically desirable for toolpath generation. In connecting the streamlines, we use a naive approach of joining streamlines with closest end points. In some cases, this may result in suboptimal connections where the resulting toolpaths are possibly shorter than what could be obtained with an optimization approach.  

In our approach, we collect user preferences only on surface quality to design support structures. However, it is possible use similar user interaction to gather high-level preferences on other qualities such as structural performance, model accuracy etc.\ and translate them to other process parameters in 3D printing similar to our approach. A natural extension to our approach would be to design variable density infill toolpaths with structural considerations.

We provide a saliency map to the user as a starting point for acquiring surface quality preferences. Although the saliency map is useful for guiding users on perceptually important features of the model, it does not provide any information on functionally important parts. In the future, our approach could be extended or complemented with a \textit{functional saliency} map to provide better prediction or guidance on the quality requirements.

%-------------------------------------------------------------------------

\section{Conclusion}
In this paper, we present an interactive tool for implicit design of support structures through high-level user preferences defined directly on the target object. Perceptual and functional impact of the support structure on the object is attenuated automatically by \one selecting the build orientation that minimize the contact with regions that are intended to have high surface quality, \two generating curvy toolpaths that conform to the shape boundaries as well as the infill patterns and \three adjusting the density of these toolpaths locally. Compared to previous methods, combination of these attributes makes our approach a practical and intuitive way to design support structures without requiring low-level knowledge on underlying support generation algorithm. 

%\section{Acknowledgments}
%ACKNOWLEDGEMENTS HERE.

%-------------------------------------------------------------------------
% bibtex
%\bibliographystyle{eg-alpha-doi}  
%\bibliography{egbibsample}        

% biblatex with biber
\printbibliography                

%-------------------------------------------------------------------------

\end{document}